\documentclass[aps,prd,preprint,tightenlines,floatfix,showpacs,groupedaddress]{revtex4}

\usepackage{subfigure}
\usepackage{graphicx}

\begin{document}
\preprint{UFIFT-HET-07-6}
\vspace{2.5in}
\date{April 30, 2007}
\vskip 2cm

\title{Does the Second Caustic Ring of Dark Matter\\ 
Cause the Monoceros Ring of Stars?}

\author{A. Natarajan$^a$ and P. Sikivie$^{a,b}$}

\affiliation{$^a$~Department of Physics, University of Florida, 
Gainesville, FL 32611, USA\\
$^b$~Theoretical Physics Division, CERN, CH-1211 Gen\`eve 23, 
Switzerland}


\begin{abstract}

Caustic rings of dark matter were predicted to exist in the 
plane of the Galaxy at radii $a_n \simeq 40~{\rm kpc}/n$ for 
$n = 1,2,3 ..$.  The recently discovered Monoceros Ring of 
stars is located near the $n=2$ caustic, prompting us to 
consider a possible connection between these two objects.  
We identify two processes through which the Monoceros Ring 
of stars may have formed. One process is the migration of 
gas to an angular velocity minimum at the caustic leading 
to enhanced star formation there. The other is the adiabatic 
deformation of star orbits as the caustic slowly grows in 
mass and radius.  The second process predicts an order 100\% 
enhancement of the density of disk stars at the location of 
the caustic ring.

\end{abstract}

\maketitle

\section{Introduction}

The Monoceros Ring is an overdensity of stars in the plane of the 
Galaxy at a galactocentric distance of approximately 20 kpc.  It was
discovered by the Sloan Digital Sky Survey (SDSS) \cite{New02,Yan03} 
and its existence was promptly confirmed by two independent 
collaborations \cite{Iba03,Roch03}.  The Monoceros Ring has been 
observed over 170 degrees in galactic longitude $l$ in the Galactic 
anti-center direction ($100^\circ \lesssim l \lesssim 270^\circ$), 
and appears to be circular.  Assuming it is a complete circle, 
the total mass in the Ring is estimated to be in the range 
$2 \cdot 10^7~-~5 \cdot 10^8~M_\odot$ in ref. \cite{Yan03}, 
and $2 \cdot 10^8~-~10^9~M_\odot$ in ref. \cite{Iba03}. The 
scale height of the Ring stars in the direction perpendicular 
to the Galactic plane is estimated to be $1.6 \pm 0.5$ kpc in 
ref. \cite{Yan03}, $0.75 \pm 0.04$ kpc in ref. \cite{Iba03} and 
$1.3 \pm 0.4$ kpc in ref. \cite{Roch03}.  The scale height in the 
direction parallel to the plane is also of order kpc.  The stars 
in the Ring move with speed approximately 220 km/s in the direction 
of galactic rotation \cite{Cran03,Yan04}.  Their velocity dispersion 
along the line of sight is small.  It was estimated to be between 
20 and 30 km/s in ref. \cite{Yan03}, and $20 \pm 4$ km/s in 
ref. \cite{Cran03}.

The most widely discussed interpretation of the Monoceros Ring 
is that it is a stream of stars resulting from the tidal disruption 
of a Galactic satellite \cite{Helm03,Mart05,Pena05}.  An alternative 
proposal is that the Ring is a manifestation of the Galactic warp 
\cite{Moma04,Moma06}.  In this paper we explore a different proposal
altogether, namely that the Monoceros Ring of stars formed as a result 
of the gravitational forces exerted by the second caustic ring of dark 
matter in the Milky Way \cite{milk}.

Caustic rings of dark matter had been predicted \cite{cr}, prior
to the discovery of the Monoceros Ring, to lie in the Galactic 
plane at radii given by the approximate law $a_n \simeq$ 40 kpc/$n$
where $n$ = 1, 2, 3 ...  Since the Monoceros Ring is located near 
the second ($n$ = 2) caustic ring of dark matter, it is natural to 
ask whether the former is a consequence of the latter.  In our 
proposal, the position of the Monoceros Ring in the Galactic 
plane and its 20 kpc radius are immediately accounted for.

Dark matter caustics are an unavoidable consequence of the fact that 
cold collisionless dark matter (CDM) lies on a 3-dim. hypersurface in 
6-dim. phase space \cite{ipser,sing,Trem,rob}.  The flow of dark 
matter in and out of the gravitational potential well of a galaxy 
necessarily produces caustics.  These caustics are of two types, 
inner and outer.  The inner caustics are rings when the angular 
momentum distribution of the dark matter is dominated by net overall 
rotation \cite{inn}.  The singularity structure of caustic rings was
discussed in detail in ref. \cite{sing}.  The ring radii were predicted 
in ref. \cite{cr} using the self-similar infall model of galactic halo 
formation \cite{FG,B}, generalized to allow the dark matter particles 
to have non-zero angular momentum \cite{STW}.  Evidence for caustic 
rings of dark matter distributed according to the prediction of the 
self-similar infall model was found in the rotation curves of exterior 
galaxies \cite{Kinn} and the rotation curve of the Milky Way \cite{milk}.

Fig.~\ref{fig1} shows the transverse section of a caustic ring.  
Fig.~\ref{fig1-a}  shows the flow of dark matter in the neighborhood 
of the caustic, whereas Fig.~\ref{fig1-b} shows the definition of its 
radius $a$ and its transverse sizes $p$ and $q$.  As was mentioned already, 
the self-similar infall model predicts that the radius $a_2$ of the second
caustic ring of dark matter is approximately 20 kpc in our galaxy.  The
transverse sizes $p$ and $q$ are not predicted by the self-similar infall 
model.  However, for reasons explained in Section II,  the expectation for
$p$ and $q$ is that they are of order 1 kpc for $n=2$. So the transverse 
sizes of the second caustic ring of dark matter are of order the
transverse sizes of the Monoceros Ring.  Moreover, for $q$ = 1 kpc, 
the caustic ring mass enclosed within the triangular shape of 
Fig.~\ref{fig1-b} is approximately $6 \cdot 10^8~M_\odot$ (see 
Section II).  Thus the total mass of the $n=2$ caustic ring of dark 
matter is of the same order of magnitude as the observed total mass 
of the Monoceros Ring.

One might ask whether it is possible to interpret the Monoceros Ring 
as a caustic in a flow of stars.  Indeed it is reasonable to assume that
the flow of dark matter that forms the $n=2$ caustic ring is accompanied
by a flow of stars occupying the same 3-dim. hypersurface in phase space.
Thus the proposal that the Monoceros Ring is a caustic in a flow of
stars would explain equally well why the Monoceros Ring is in the 
Galactic plane and why its radius is 20 kpc.  However, the proposal 
runs into difficulties.  The first difficulty is that the self-similar 
infall model, which predicts the 20 kpc radius of the $n=2$ caustic 
ring, also predicts that the matter in that caustic moves with a 
velocity of approximately 515 km/s in the direction of Galactic 
rotation (see Section II).  This is inconsistent with the 220 km/s 
observed velocity of Monoceros Ring stars in the direction of galactic 
rotation.  A second difficulty is that the matter in the caustic 
ring has an effective radial velocity dispersion of order 100 km/s, 
whereas the radial velocity dispersion of Monoceros Ring stars is 
only of order 20 km/s.  Thus the proposal that the Monoceros Ring 
is a caustic of stars appears untenable.  As mentioned already, the
proposal we explore instead is that the Monoceros Ring is caused by 
the gravitational field of the $n=2$ caustic ring of dark matter.

We will find that there are two apparently viable mechanisms 
by which the $n=2$ caustic ring of dark matter may cause the 
Monoceros Ring of stars.  The first mechanism is enhanced 
star formation at the caustic radius because viscous forces 
on the gas in the neighborhood of the caustic drive it 
towards $r = a$.  The second mechanism is the adiabatic 
deformation of the orbits of ordinary disk stars by the 
caustic ring of dark matter.  We will show that the increase 
in star density at the location of the caustic as a result of 
this second mechanism is of order 100\%.

The paper is organized as follows: In Section II, we give 
a self-contained and hopefully pedagogical description of 
the properties of caustic rings of dark matter, including the 
predictions of the self-similar infall model for the parameters 
that characterize the caustic rings, the observational evidence 
for caustic rings of dark matter, and the expected properties  
of the $n = 2$ ring.  In Section III, we describe the gravitational 
field of a caustic ring of dark matter and the perturbation it 
causes to the galactic rotation curve.  We then analyze two 
mechanisms by which the second caustic ring of dark matter 
may cause the Monoceros Ring: the migration of gas to a sharp 
minimum in the angular velocity at $r = a$ implying enhanced 
star formation there, and the adiabatic deformation of ordinary 
disk star orbits by the slowly growing caustic.  Section IV 
provides concluding remarks.  Axial symmetry is assumed 
throughout unless stated otherwise.

\section{Caustic ring properties}

Cold collisionless dark matter (CDM) particles lie on a 3-dim. 
hypersurface in phase space \cite{ipser,sing,Trem}.  We refer
to this 3-dim. hypersurface as the ``phase space sheet".  At the 
location of a galactic halo, the phase space sheet is wound up 
so as to cover physical space multiple times.  This phase space
structure implies that the velocity distribution of CDM particles 
is everywhere discrete and that there are surfaces in physical space, 
called caustics, where the density of dark matter particles is very 
large.  Discrete flows and caustics are a robust prediction of cold 
dark matter cosmology \cite{rob}.  The reader may wish to consult 
ref. \cite{rob} for background information and a list of references.

Galactic halos have outer caustics and inner caustics.  The outer 
caustics are a set of simple fold ($A_2$) catastrophes located on
topological spheres surrounding the galaxy at radii of order hundreds 
of kpc.  Our focus, however, is on the inner caustics, which are much 
closer to the galactic center.  The physical shape and catastrophe 
structure of inner caustics depend on the angular momentum distribution 
of CDM particles falling onto the halo \cite{inn}.  If that angular
momentum distribution is dominated by net overall rotation, the inner 
caustics are a set of ring-like closed tubes, called ``caustic rings", 
in or near the galactic plane.  In cross-section each tube is a section 
of the elliptic umbilic ($D_{-4}$) catastrophe \cite{sing}.  

In this section we give a detailed desciption of caustic rings in the 
limit of axial symmetry and where their cross-sectional sizes, $p$ and 
$q$, are much smaller than their radius $a$.  Under these assumptions, 
the distribution of CDM particles in the vicinity of the caustic ring 
is determined in terms of a relatively small number of parameters, which 
we identify.  Next, we summarize the predictions of the self-similar infall 
model for the caustic ring properties.  We briefly review the evidence for
caustic rings of dark matter, at the radii predicted by the self-similar 
infall model, in the Milky Way and in other isolated spiral galaxies.  
Finally, we list the expected properties of the second caustic ring of 
dark matter in our galaxy.

\subsection{Catastrophe structure}

The caustic ring singularity was analyzed in ref.~\cite{sing} which
the reader may wish to consult for details.  In the limit of axial 
symmetry and where the transverse sizes, $p$ and $q$, of a caustic 
ring are much smaller than its radius $a$, the distribution of CDM 
in the vicinity of the caustic is given by the particle positions
\begin{eqnarray}
z(\alpha, \tau) &=& b \alpha \tau  \nonumber\\
\rho(\alpha, \tau) &=& a + {1 \over 2} u (\tau - \tau_0)^2
- {1 \over 2} s \alpha^2~~~~\ .
\label{crfl}
\end{eqnarray}
We use cylindrical coordinates $(z, \rho, \phi)$ for position 
in physical space.  Eqs.~(\ref{crfl}) give particle positions 
at a particular time, say $t=0$.  The particles are labeled by 
parameters $(\alpha, \tau)$.  $\alpha \equiv {\pi \over 2} - \theta$ 
where $\theta$ is the polar angle of the particle at the time of its 
last turnaround.  $\tau$ is the time when the particle crosses 
the $z = 0$ plane.  $t - \tau$ can be thought of as the age of
the particle.  The particles labeled $(\alpha, \tau)$ form a 
circle of radius $\rho(\alpha, \tau)$ at a height $z(\alpha, \tau)$ 
above the $z = 0$ plane.  

$b$, $a$, $u$, $\tau_0$ and $s$ are constants characterizing the caustic
ring.  Each has physical meaning.  See ref. \cite{sing} for more precise
descriptions than we give here. $a$ is the radius of the caustic ring.
$|\tau_0|$ is of order the time a constituent particle spends in the
caustic.  $b$ is of order the speed of the particles in the caustic. 
$u$ is of order their centrifugal acceleration. $s$ characterizes the
$\alpha$-dependence of specific angular momentum near the equator 
($\alpha = 0$).

Fig.~\ref{fig1} describes the caustic ring cross-section.  Fig.~\ref{fig1-a}
plots $(\rho(\alpha, \tau), z(\alpha, \tau))$ for continuous $\tau$, 
and discrete values of $\alpha$.  The lines in Fig.~\ref{fig1-a} are 
the trajectories of the particles forming the flow, except that 
positions are plotted as a function of age, whereas for ordinary 
trajectories position is plotted as a function of time.  Let 
us call the lines of Fig.~\ref{fig1-a} ``age trajectories''.  
Fig.~\ref{fig1-a} shows that particle density diverges on a 
closed line which has the shape of a isosceles triangle, but 
with cusps instead of angles.  We call that shape a``tricusp''.  
The location of the tricusp is shown in Fig.~\ref{fig1-b} for 
the flow of Fig.~\ref{fig1-a}.  It is the envelope of the age 
trajectories.  There are four flows everywhere inside the tricusp 
and two flows everywhere outside.  The caustic, i.e. the surface 
where the density diverges, lies at the boundary between the region 
with four flows and the region with two flows.

The physical space density is given by
\begin{equation}
d(\rho, z) = {1 \over \rho} \sum_{j=1}^{N(\rho,z)}
{dM \over d\Omega d\tau}(\alpha,\tau)
{\cos\alpha \over |D(\alpha, \tau)|}
\bigg|_{(\alpha_j(\rho,z), \tau_j(\rho,z))}
\label{den}
\end{equation}
where $\alpha_j(\rho,z)$ and $\tau_j(\rho,z)$, with 
$j = 1~...~N(\rho,z)$, are the solutions of $\rho(\alpha,\tau) = \rho$
and $z(\alpha,\tau) = z$. $N(\rho,z)$ is the number of flows at 
position $(\rho,z)$; thus, $N=4$ inside the tricusp and $N=2$ 
outside.  $D(\alpha,\tau)$ is the Jacobian determinant of the 
map $(\alpha,\tau) \rightarrow (\rho,z)$:
\begin{equation}
D(\alpha,\tau) \equiv
\det\left({\partial(\rho,z) \over \partial (\alpha,\tau)}\right)
= - b [u \tau (\tau - \tau_0) + s \alpha^2]~~~\ .
\label{D}
\end{equation}
${dM \over d\Omega d\tau} = {dM \over 2 \pi \cos\alpha d\alpha d\tau}$
is the mass falling in per unit solid angle and unit time.  The tricusp 
perimeter is the locus of points $(\rho(\alpha, \tau), z(\alpha, \tau))$ 
for which $D(\alpha, \tau) = 0$.  We call $p$ and $q$ the sizes of the
tricusp in the $\rho$ and $z$ directions respectively; see Fig.~\ref{fig1-b}.  
They are given by 
\begin{equation}
p = {1 \over 2} u \tau_0^2~~,~~~ 
q = {\sqrt{27} \over 4} {b \over \sqrt{us}}~p~~~\ .
\label{pq}
\end{equation}
One of the cusps of the tricusp points away from the galactic center.

Consider the 3-dim. space spanned by the two physical coordinates 
$\rho$ and $z$ plus the constant $\tau_0$.  $\tau_0$ can be positive 
or negative.  In $(\rho, z, \tau_0)$ space, the tricusp is perpendicular
to the $\tau_0$ axis with one cusp pointing in the positive $\rho$
direction.  As $\tau_0$ varies from negative to positive values, 
the size of the tricusp, which varies as $\tau_0^2$, shrinks to zero
and then increases again.  The structure in the neighborhood of 
$\tau_0 = 0$ is the (full) elliptic umbilic catastrophe ($D_{-4}$), 
one of the elementary catastrophes in three dimensions.  Thus the 
tricusp, i.e. the cross-section of a caustic ring, is a section of 
the elliptic umbilic.

If the $z$-axis is rescaled relative to the $\rho$-axis so 
as to make the tricusp equilateral, the tricusp has a $Z_3$
symmetry \cite{sing} consisting of rotations by multiples of 
${2 \pi \over 3}$ about the point of coordinates
$(\rho_c,z_c) = (a + p/4, 0)$.  This point may thus be called 
the center of the tricusp.  It is indicated by a star in Fig.~\ref{fig1-b}.
As an example of how the above formalism is to be used, let us 
derive a formula for the density $d_c$ at the center of the
tricusp.  Using Eqs. (\ref{crfl}), one finds that the four flows 
there have parameter values $(\alpha, \tau) = 
(0, {3 \over 2} \tau_0),~(0, {1 \over 2} \tau_0),~
(+\sqrt{3p \over 2s}, 0)$ and $(-\sqrt{3p \over 2s}, 0)$.  
The corresponding values of the Jacobian determinant are 
$D = - {3 \over 2} bp,~{1 \over 2} bp,~ - {3 \over 2} bp$ and 
$ - {3 \over 2} bp$.  Inserting these in Eq.~(\ref{den}), one 
finds
\begin{equation}
d_c = {4 \over a b p} {dM \over d\Omega d\tau}~~~\ .
\label{cenden}
\end{equation}
In obtaining this result, we approximated $\rho$ by $a$ and
$\cos \alpha$ by 1 in Eq.~(\ref{den}), and neglected the
$\alpha$-dependence of ${dM \over d\Omega d\tau}$.  These 
approximations are appropriate since we assume that $p$ and 
$q$ are small relative to $a$.  For the mass per unit length 
enclosed within the tricusp, we find by numerical integration
\begin{equation}
\lambda = 0.6~ p q d_c~~~\ .
\label{mol}
\end{equation}
Note that some of the mass associated with the caustic ring lies 
outside the tricusp.  In particular, when $p = q = 0$, $\lambda = 0$ 
but there is still a caustic ring and an associated overdensity.  In 
that limit, the tricusp has collapsed to a point, and the particle 
density diverges as the inverse distance to that point \cite{sing}.

Finally, we describe the velocities of the particles that constitute
the caustic ring.  Let us write the velocity of particle labeled
$(\alpha, \tau)$ as  $\vec{v}(\alpha, \tau) = 
v_\phi (\alpha, \tau) \hat{\phi} + v_\rho (\alpha, \tau) \hat{\rho}
+ v_z (\alpha, \tau) \hat{z}$.  For $p,q \ll a$, all the particles 
in the caustic have the same speed $v$ because they have come to 
the caustic from a small region (near the equator) of the turnaround
sphere, along neighboring trajectories.  The main component of velocity 
is in the $\hat{\phi}$ direction: $v_\phi \simeq v$.  The speed $v$ is 
related to the centrifugal acceleration $u$ by 
\begin{equation}
u = {v^2 \over a}~~~\ .
\label{ca}
\end{equation}
In the case of a stationary flow, the velocity components in the 
$\hat{\rho}$ and $\hat{z}$ directions are 
\begin{equation}
v_\rho = - {\partial \rho \over \partial \tau} = - u (\tau - \tau_0)~~,~~~
v_z = - {\partial z \over \partial \tau} = - b \alpha~~~\ .
\label{tranvel}
\end{equation}
Here we use the fact that, in case of stationary flow, the 
particle positions are functions only of their age $t - \tau$.
For the four flows within the tricusp, $v_\rho$ and $v_z$ are of 
order $u \tau_0 = v \sqrt{2p \over a}$.  For the four flows at 
the center, for example, $(v_\rho, v_z) = 
(-{1 \over 2} u \tau_0, 0),~(+{1 \over 2} u \tau_0, 0),~
(u \tau_0, - {1 \over 2} \sqrt{3 \over \zeta} u \tau_0),~
(u \tau_0, + {1 \over 2} \sqrt{3 \over \zeta} u \tau_0)$, 
where $\zeta \equiv {us \over b^2} = {16 p^2 \over 27 q^2}$.

Caustic rings grow in mass and radius on cosmological time scales, 
implying that the flow is not exactly stationary.  However, provided 
one has information on the manner in which the growth occurs, the 
time dependence is easily included.  For example, if the growth in 
size is tantamount to an increase in the caustic radius at the rate 
$\dot{a}$, the velocities are still as described in the previous 
paragraph except $v_\rho$ is replaced with $v_\rho + \dot{a}$.  If 
the growth is tantamount to expansion in all directions by a scale 
factor $R(t)$, $v_\rho$ is replaced by  $v_\rho + {\dot{R} \over R} \rho$ 
and $v_z$ by $v_z + {\dot{R} \over R} z$.

Thus far, we have given a detailed description of a caustic 
ring in the limit of axial symmetry and where the transverse
sizes $p$ and $q$ of the ring are much smaller than its radius 
$a$.  The description is in terms of a small number of parameters:
$a,~b,~u,~\tau_0,~s$ and ${dM \over d\Omega d\tau}$.  We now 
turn to the self-similar model of galactic halo formation 
to obtain estimates of these parameters for the caustic 
rings in actual halos.

\subsection{Predictions of the self-similar infall model}

The self-similar model of galactic halo formation \cite{FG,B,STW}
assumes that the entire halo phase space distribution is unchanged 
in time except for a rescaling of all lengths by a scale factor 
$R(t)$, and all velocities by ${R(t) \over t}$ where $t$ is cosmic 
time.  Physical space densities scale as ${1 \over t^2}$.  A 
spherically symmetric overdensity in an Einstein-de Sitter universe 
($\Omega_{\rm matter} = 1$) has self-similar evolution provided its 
initial profile is a power law \cite{FG,B}
\begin{equation}
{\delta M_i \over M_i} = \left({M_0 \over M_i}\right)^\epsilon~~~\ ,
\label{prof}
\end{equation}
where $M_i$ is the mass interior to initial radius $r_i$, $\delta M_i$ 
is the corresponding extra mass, and $\epsilon$ is a parameter with the 
a-priori range $0 \leq \epsilon \leq 1$.  The scale factor $R(t)$ is 
proportional to $t^{{2 \over 3} + {2 \over 9 \epsilon}}$.  The rotation 
curve is flat at small $r$ provided 
$0 \leq \epsilon \leq {2 \over 3}$ \cite{FG}.  In an average sense,
$\epsilon$ is related to the slope of the power spectrum of density 
perturbations on galactic scales \cite{Dor}.  The standard CDM power 
spectrum implies $\epsilon$ is in the range 0.2 to 0.35 \cite{STW}.

The accelerated expansion \cite{acc} of the universe is not 
consistent with Einstein-de Sitter cosmology, nor therefore 
with strict self-similarity.  This is not a serious shortcoming 
of the self-similar model, however, because galactic halos were 
formed for the most part long before the onset of accelerated 
expansion, when the universe was accurately described by 
Einstein-de Sitter cosmology.   

The original spherically symmetric self-similar model \cite{FG,B} 
assumes radial orbits for all the particles.  Whereas this approximation 
is reasonable when describing the outer parts of a galactic halo, it is 
inadequate for the inner parts.  Specifically, in the approximation of
radial orbits, all particles pass through the galactic center each 
time they fall in and out of the galaxy, causing the density to 
diverge at the center as ${1 \over r^2}$.  The halo contribution 
to the rotation curve then approaches a constant as $r \rightarrow 0$.
In actual galaxies, the central parts are dominated by baryons, and the 
halo contribution to the rotation curve goes to zero at the center.  

However, the self-similar model can be generalized \cite{STW} to allow 
the dark matter particles to have angular momentum.  Self-similarity is 
maintained provided the specific angular momentum distribution on the 
turnaround sphere at time $t$ is of the form:
\begin{equation}
\vec{\ell}(\hat{r},t) = \vec{j}(\hat{r}) {R^2(t) \over t}~~~\ ,
\label{am}
\end{equation}
where $R(t)$ is the turnaround radius and $\vec{j}(\hat{r})$ is a 
dimensionless and time-independent angular momentum distribution.  
When angular momentum is included, particle orbits avoid the galactic
center, the inner parts of the halo are depleted, and the halo 
contribution to the rotation curve goes to zero at $r = 0$, as 
it should.  In the Milky Way, approximately half of the rotation 
velocity squared at our location is due to dark matter, and half 
is due to baryonic matter.  This determines the average magnitude 
$\bar{j}$ of the rescaled angular momentum distribution 
$\vec{j}(\hat{r})$ to be of order $\bar{j} \sim 0.2$ for the Milky 
Way halo \cite{STW}.

The caustic rings of dark matter occur where the particles with 
the most angular momentum are at their distance of closest approach 
to the galactic center \cite{sing,inn}.  In the spherically
symmetric model, the orbits are radial and hence the distance of 
closest approach vanishes for all particles.  In that case all the 
caustic rings collapse to a single caustic point at the center of 
spherical symmetry.  In the more realistic self-similar model with 
angular momentum, the caustic rings have finite radii.  The radii 
are predicted in terms of the maximum value $j_{\rm max}$ of the 
dimensionless angular momentum distribution $\vec{j}(\hat{r})$
\cite{cr}:
\begin{equation}
\{a_n : n = 1, 2, 3, 4, 5 ...\} \simeq ~{40~{\rm kpc} \over n}~ 
\left({v_{\rm rot} \over 220~{\rm km/s}}\right)
\left({j_{\rm max} \over 0.25}\right)
\label{crr}
\end{equation}
where $v_{\rm rot}$ is the rotation velocity of the galaxy.  
Eq.~(\ref{crr}) is for the particular case $\epsilon = 0.3$. 
However, the $a_n \propto {1 \over n}$ approximate behaviour 
holds for all $\epsilon$ in the relevant range 
$0.2 \leq \epsilon \leq 0.35$, so that a change in $\epsilon$ 
can be compensated for, as far as the $a_n$ values are concerned, by 
a change in $j_{\rm max}$.  We will assume $\epsilon = 0.3$ henceforth.  
The self-similar model does not predict $j_{\rm max}$ and allows
a different $j_{\rm max}$ value for each galactic halo.

The relationship between $\bar{j}$ and $j_{\rm max}$ depends, of 
course, on the $\vec{j}(\hat{r})$ distribution.  Let us assume the
simplest distribution consistent with net overall rotation and axial 
symmetry:  $\vec{j}(\hat{r}) = j_{\rm max} \hat{\phi} \cos\alpha$,
i.e. that the turnaround sphere is initially rigidly rotating.  In 
that case, $\bar{j} = {\pi \over 4} j_{\rm max}$.  Thus, for the 
Milky Way halo, the estimate $\bar{j} \sim 0.2$ implies 
$j_{\rm max} \sim 0.25$.  

In summary so far, the self similar infall model with axial symmetry 
and net overall rotation predicts caustic rings in the galactic plane 
at the radii specified in Eq.~(\ref{crr}).  It also predicts the speeds 
$v_n$ of the dark matter particles forming the caustic rings and the
infall rates ${dM \over d\Omega d\tau}|_n$.  The $v_n$ are approximately 
$n$-independent:
\begin{equation}
v_n \simeq~515~{\rm km \over s}
\left({v_{\rm rot} \over 220~{\rm km/s}}\right)~~~\ .
\label{spee}
\end{equation}
The infall rates are given by
\begin{equation}
{dM \over d\Omega d\tau}{1 \over v}\Bigg|_n
= f_n {v_{\rm rot}^2 \over 4 \pi G}
\label{infr}
\end{equation}
with
\begin{equation}
\{f_n : n = 1,2,3,4,5, ... \} \simeq
(11,~4.6,~2.9,~2.1,~1.7,~...)\times 10^{-2}~~~\ .
\label{fs}
\end{equation}
Eqs.~(\ref{infr}) and (\ref{fs}) provide the prefactor which
appears in the formulas for the density [Eqs.~(\ref{den})
and (\ref{cenden})], up to the ratio ${v \over b}$, which is
of order one but which may differ from one by a factor two or
so.  The transverse sizes, $p$ and $q$, of the caustic rings 
and the ratio ${v \over b}$ depend on relatively more subtle 
properties of the velocity distribution at last turnaround 
\cite{sing}, and are not predicted by the self-similar infall 
model.

Observational evidence has been found in support of Eq.~(\ref{crr}) 
in the Milky Way and other spiral galaxies.  We briefly describe 
this evidence now.

\subsection{Summary of observational evidence}

Caustic rings of dark matter in or near the galactic plane cause bumps 
in the galactic rotation curve.  In ref.~\cite{Kinn} a set of 32 extended
and well-measured rotation curves was analyzed to test the hypothesis that 
some of their bumps are caused by caustic rings of dark matter at the radii 
given by Eq.~(\ref{crr}).  For each rotation curve, the radial coordinate
$r$ was rescaled according to 
\begin{equation}
r~\rightarrow~\tilde {r} = r
\left({220~{\rm km/s} \over v_{\rm rot}}\right)~~~\ ,
\label{resc}
\end{equation}
where $v_{\rm rot}$ is the rotation velocity implied by the curve. The 
rotation curves were then co-added.  The combined rotation curve shows 
peaks at $\tilde{r} \simeq 20$ and 40 kpc, with a significance of 
2.6 $\sigma$ and 3.0 $\sigma$ respectively.  The result suggests 
not only the existence of caustic rings of dark matter at the 
radii given by Eq.~(\ref{crr}), but also that the distribution of 
$j_{\rm max}$ values is peaked near $j_{\rm max} \simeq 0.27$.  

The Milky Way rotation curve has a series of sharp rises at radii 
which agree at the 3\% level \cite{milk} with the caustic ring 
radii given in Eq.~(\ref{crr}).  The inner North Galactic rotation 
curve of ref. \cite{Clem} has ten rises between $r$ = 3 and 8.5 kpc, 
which may be identified with caustic rings $n = 5,~6~...~14$.  
The rises are ``sharp" in the sense that they start and end with
discontinuities (kinks) in the slope of the rotation curve.  Kinks 
in the rotation curve are predicted by caustic rings because the 
dark matter density diverges at the caustic.  The outer Milky Way 
rotation curve is much less well measured.  Nonetheless it has a 
prominent rise near $r$ = 13 kpc which may be identified with the 
$n = 3$ caustic ring.  Finally, the IRAS map of the Galactic plane 
in the direction of galactic coordinates $(l,b) = (80^\circ, 0^\circ)$ 
has a triangular feature whose position and appearance is consistent 
with the imprint of the caustic ring of dark matter nearest to us 
($n = 5$) upon the gas and dust in the disk.

\subsection{Expectations for the $n=2$ caustic ring}

Eq.~(\ref{crr}) predicts that the $n=2$ caustic ring has radius
$a_2 \simeq 20$ kpc.  Eq.~(\ref{spee}) predicts that the speed 
of the particles constituting the ring is approximately 515 km/s.  
The transverse sizes $p_2$ and $q_2$ are not predicted.  However, 
it is reasonable to expect that the $n=2$ caustic ring has properties 
similar to caustic rings $n=3$ and $n=5$ to 14, for which we have 
the observational evidence mentioned in the previous subsection. 
The $p$ values of those rings can be read off from the widths of 
the corresponding rises in the Milky Way rotation curve \cite{milk}.  
For those ``observed" rings, one finds that $p/a$ ranges from 0.015 
to 0.1, with an average value of 0.05.  We thus expect $p_2$ to be 
of order 1 kpc, and very likely between 0.3 and 2 kpc.  The rises 
in the rotation curve do not inform us about the values of $q$.  
However, the triangular feature in the IRAS map provides values 
of both $p$ and $q$ for the $n = 5$ caustic ring. In that case 
${q \over p} \simeq 1.5$. We therefore expect $q_2$ to be of 
order 1 kpc as well.  

The prediction for the dark matter density at the central point 
of the tricusp is obtained by combining Eqs.~(\ref{cenden}), 
(\ref{infr}) and (\ref{fs}):
\begin{equation}
d_{c,2} = 8 \cdot 10^{-3}~{M_\odot \over {\rm pc}^3}~
\left({v_2 \over b_2}\right)~
\left({{\rm kpc} \over p_2}\right)~~~\ .
\label{dc2}
\end{equation}
Using Eq.~(\ref{mol}), we obtain the predicted dark matter
mass per unit length enclosed within the tricusp 
\begin{equation}
\lambda_2 = 5 \cdot 10^6~{M_\odot \over {\rm kpc}}~
\left({v_2 \over b_2}\right)~
\left({q_2 \over {\rm kpc}}\right)~~~\ .
\label{mol2}
\end{equation}
Since $b_2$ is of order $v_2$, the total mass $2 \pi a_2 \lambda_2$
enclosed within the tricusp is of order $6 \cdot 10^8~M_\odot$.  
Finally, from Eq.~(\ref{tranvel}) we learn that the transverse 
velocity components $v_\rho$ and $v_z$ of the particles within 
the tricusp are of order (515 km/s)$\sqrt{2 p_2 \over a_2} \sim 160$
km/s.

\section{Effect of a caustic ring of dark matter upon baryonic 
matter in the disk} 

In this section, we first describe the gravitational field of 
a caustic ring of dark matter and its effect on the Galactic 
rotation curve, assuming that the caustic ring lies in the 
plane of the disk.  Next we identify two possible mechanisms
by which a ring of stars may form due to the presence of a 
caustic ring of dark matter.  The first mechanism is the 
migration of gas to the caustic ring radius, leading to 
enhanced star formation there.  The second is the adiabatic 
deformation of star orbits as the caustic ring slowly grows in 
size.

\subsection{Gravitational field of a caustic ring}

As before, we assume that the transverse dimensions, $p$ and $q$, 
of the caustic ring are small compared to its radius $a$.  In that 
limit, when calculating the gravitational field $\vec{g}$ of the 
caustic ring at a distance of order $p$ or $q$ from the ring, we
may neglect its curvature and pretend that it is a straight tube.
Then 
\begin{equation}
\vec{g}_c(\rho,z) = - 2 G \int d\rho^\prime dz^\prime~ 
d(\rho^\prime, z^\prime) 
{(\rho - \rho^\prime, z - z^\prime) \over 
(\rho - \rho^\prime)^2 + (z - z^\prime)^2}~~~\ .
\label{gcg}
\end{equation}
Using Eq.~(\ref{den}), changing variables 
$(\rho^\prime, z^\prime) \rightarrow (\alpha, \tau)$, 
neglecting the $(\alpha, \tau)$ dependence of 
${dM \over d\Omega d\tau} (\alpha, \tau)$ over the 
size of the caustic, and approximating $\cos \alpha \simeq 1$, 
we obtain
\begin{equation}
\vec{g}_c(\rho, z) = - {2 G \over \rho} {dM \over d\Omega d\tau}
\int d\alpha d\tau  
{(\rho - \rho(\alpha, \tau), z - z(\alpha, \tau)) \over
(\rho - \rho(\alpha, \tau))^2 + (z - z(\alpha, \tau))^2}~~~\ .
\label{gcat}
\end{equation}
Eqs.~(\ref{crfl}) provide the functions $\rho(\alpha, \tau)$
and $z(\alpha,\tau)$.  The integral on the RHS of Eq.~(\ref{gcat})
may be evaluated numerically for all $\rho$ and $z$.  In the 
galactic plane 
\begin{equation}
\vec{g}_c(r,0) = - {8 \pi G \over r b} {dM \over d\Omega d\tau}
I(\zeta, {r-a \over p}) \hat{r}
\label{gcr0}
\end{equation}
with $\zeta = {s u \over b^2}$ (as before) and 
\begin{equation}
I(\zeta,X) \equiv {1 \over 2 \pi} \int dA dT
{X - (T-1)^2 + \zeta A^2 \over 
[X - (T-1)^2 + \zeta A^2]^2 + 4 A^2 T^2}~~~\ ,
\label{IX}
\end{equation}
where $X = {r-a \over p}$.  For $\zeta = 1$ one can do the 
integral by analytical methods, with the result
\begin{eqnarray}
I(1,X) &=& - {1 \over 2}~~~~~~~~~~~~~~~~~~{\rm for}~~X<0~~~\nonumber\\
&=& - {1 \over 2} + \sqrt{X}~~~~~~~~~{\rm for}~~0<X<1~~~\nonumber\\
&=& + {1 \over 2}~~~~~~~~~~~~~~~~~~{\rm for}~~X>1~~~~\ .
\label{I1X}
\end{eqnarray}
A plot of $I(1,X)$ is shown in Fig.~\ref{fig2-a}.  $I(1,X)$ has 
an upward kink at $X=0$ and a downward kink at $X=1$.  The kinks
are due to the divergent behaviour of the dark matter density 
in the $z=0$ plane at $r=a$ and $r=a+p$.  The shape of $I(\zeta,X)$ 
with $\zeta \neq 1$ is similar to that of $I(1,X)$.  The amount 
$\Delta I(\zeta)$ by which $I(\zeta,X)$ rises between $X=0$ and 
$X=1$ depends on $\zeta$.  A plot of $\Delta I(\zeta)$ is given in 
ref.~\cite{sing}.  $\Delta I(\zeta)$ is near one for $\zeta$
of order one, and $\Delta I(1) = 1$.

Substituting Eq.~(\ref{infr}) into Eq.~(\ref{gcr0}), we obtain:
\begin{equation}
\vec{g}_c(r,0) = - 2 f {v_{\rm rot}^2 \over r} {v \over b}
I(\zeta, {r-a \over p}) \hat{r}~~~~~\ .
\label{gcrot}
\end{equation}
Since ${v_{\rm rot}^2 \over r}$ is the gravitational field of the 
galaxy as a whole, and ${v \over b}$ and $I(\zeta, {r-a \over p})$ are
both of order one whereas $f$ is of order a few \%, Eq.~(\ref{gcrot}) 
implies that the caustic ring of dark matter causes only a small 
perturbation on the local gravitational field.

Consider a smooth halo which produces a flat rotation curve with 
rotation velocity $v_{\rm rot}$. If we add to the halo a caustic
ring of dark matter of radius a, the gravitational field near 
$r=a$ would be $\vec{g}(\vec{r}) = - g(r) \hat{r}$ with 
\begin{equation}
g(r) = {v_{\rm rot}^2 \over r} [1 + 2 f {v \over b} 
I(\zeta, {r-a \over p})]~~~\ .
\label{incg}
\end{equation}
However, this is not what we want to do.  Indeed, the caustic ring 
of dark matter is not an {\it addition} to the smooth halo.  Rather, 
it is the outcome of {\it redistributing} dark matter already 
present in the halo.  One can see that it is incorrect merely 
to add caustic rings of dark matter to a smooth halo by noting 
that this would cause the rotation curve to rise.  The rotation 
velocity squared would rise by the relative amount 
$2 f_n {v_n \over b_n} \Delta I(\zeta_n)$ at the $n$th ring, 
for each ring.  Instead, the redistribution of dark matter 
into caustic rings should be such that the rotation curve 
remains flat on average.  The gravitational field in the 
vicinity of a caustic ring is therefore
of the form
\begin{equation}
g(r) = {v_{\rm rot}^2 \over r}\left(1 + 2 f {v \over b}
[I(\zeta, {r-a \over p}) + H(r)]\right)
\label{corrgr}
\end{equation}
where $H(r)$ is a smooth function which steps down by the amount
$\Delta I (\zeta)$ as $r$ increases from $r \ll a$ to $r \gg a$;
for example, 
$H(r) = - {1 \over 2} \Delta I(\zeta) 
\tanh ({r-a \over p^\prime})$. $p^\prime$ is expected to be of 
order $a$.  Let us define
\begin{equation}
J(r - a) \equiv {v \over b} [I(\zeta, {r-a \over p}) + H(r)]~~~\ .
\label{Jr}
\end{equation}
The gravitational field in the galactic plane near a caustic 
ring of dark matter is then
\begin{equation}
g(r) = {v_{\rm rot}^2 \over r} [1 + 2 f J(r-a)]~~~\ .
\label{fingr}
\end{equation}
Fig.~\ref{fig2-b} shows what the function $J(r-a)$ looks 
like qualitatively.

\subsection{Effect on gas}

Gas in circular orbits in the galactic plane moves with 
angular velocity
\begin{equation}
\Omega(r) = \sqrt{g(r) \over r}~~~~\ .
\label{angvel}
\end{equation}
Unless $\Omega$ is $r$-independent (rigid rotation) there 
is shear in the velocity field.  The viscous forces which 
result from this shear cause radial motion of the gas, as 
discussed by Lynden-Bell and Pringle \cite{LBP}.  For the 
sake of completeness, we repeat here relevant considerations 
from ref. \cite{LBP}.

Consider a cylinder of radius $r$.  The viscous force per unit 
area across the cylinder is $\rho \nu r {d \Omega \over dr}$
where $\rho$ is the gas density and $\nu$ its viscosity.  Hence, 
the viscous torque across the cylinder is 
\begin{equation}
\tau = 2 \pi r \int dz~(\rho \nu r {d \Omega \over dr})~r
= 2 \pi \nu \sigma r^3 {d \Omega \over dr}
\label{tor}
\end{equation}
where $\sigma = \int dz~\rho$ is the mass per unit surface 
of the gas.  Consider an annulus of width $\delta r$.  Its 
mass is $\delta m = 2 \pi r \delta r \sigma$.  The torque 
on the annulus is 
\begin{equation}
\tau(r + \delta r) - \tau(r) = 
\delta r {d \over dr}(2 \pi \nu \sigma r^3 {d \Omega \over dr})
= {d \over dt} \delta L
\label{anntor}
\end{equation}
where $\delta L = \delta m \Omega(r) r^2$ is its angular 
momentum, and ${d \over dt} = {\partial \over \partial t} + 
v_r {\partial \over \partial r}$ is the time derivative 
``following the motion".  We have ${d~\delta m \over dt} = 0$.  
Also, ${\partial \over \partial t} [\Omega(r) r^2] = 0$ 
because the gravitational field is determined by the 
mass distribution of the Galaxy as a whole and only 
negligibly altered by the motion of the gas.  We thus 
obtain
\begin{equation}
v_r = {1 \over \sigma r}
{{d \over dr} (\nu \sigma r^3 {d \Omega \over dr}) \over 
{d \over dr} (\Omega r^2)}
\label{vr}
\end{equation}
for the radial velocity of the gas.  The gas obeys the 
continuity equation
\begin{equation}
{\partial \sigma \over \partial t} + 
{1 \over r} {\partial \over \partial r} (r v_r \sigma) = 0~~~\ .
\label{con}
\end{equation}
Eq.~(\ref{vr}) neglects the back-reaction pressure from 
the accumulation of gas at particular radii.

Let us assume that the gas is distributed uniformly 
$({d \sigma \over dr} = {d \nu \over dr} = 0$) to 
start with, and use Eq.~({\ref{vr}) to determine in 
which direction it is driven by the viscous forces.  
In a smooth halo with flat rotation curve, 
$\Omega = {v_{\rm rot} \over r}$ and hence 
$v_r = - {\nu \over r}$.  If a caustic ring 
of dark matter lies in the disk, we have instead 
in the neighborhood of the ring
\begin{equation}
\Omega(r) = {v_{\rm rot} \over r} [1 + 2 f J(r-a)]^{1 \over 2}
= {v_{\rm rot} \over r} [1 + f J(r-a) + 0(f^2)]~~~~\ .
\label{crom}
\end{equation}
Inserting Eq.~(\ref{crom}) into Eq.~(\ref{vr}), setting 
${d \sigma \over dr} = {d \nu \over dr} = 0$, and neglecting 
terms of order $f^2$, we obtain
\begin{equation}
v_r = - {\nu \over r}~~
{1 + f J(r-a) - r f {dJ \over dr} - r^2 f {d^2 J \over dr^2} \over 
1 + f J(r-a) + r f {dJ \over dr}}~~~ \ .
\label{crvr}
\end{equation}
To obtain the qualitative behaviour, we neglect 
$f J(r-a)$, $r f {dH \over dr}$ and $r^2 f {d^2 H \over dr^2}$
vs. 1, because such terms are all of order f.  Then
\begin{equation}
v_r \simeq - {\nu \over r}~~ 
{1 - f r {dI \over dr} - f r^2 {d^2 I \over dr^2} \over 
1 + f r {dI \over dr}}~~~\ .
\label{avr}
\end{equation}
For $r < a$ and $r > a+p$, ${d I \over dr} =0$ and hence 
$v_r  = - {\nu \over r}$ as in the absence of caustic.  For
$a < r < a+p$, 
\begin{equation}
v_r \simeq - {\nu \over a}~~ 
{1 - {fa \over 2p} {1 \over \sqrt{X}} + 
{f a^2 \over 4 p^2} {1 \over X^{3 \over 2}} \over 
1 + {fa \over 2p} {1 \over \sqrt{X}}}~~~\ ,
\label{vrnf}
\end{equation}
where $X = {r-a \over p}$, as before.  Since ${fa \over 2p} = 0(1)$
whereas ${fa^2 \over 4 p^2} = 0(10)$, the numerator on the RHS of 
Eq.~(\ref{vrnf}) is always dominated by the last term.  Hence
\begin{equation}
v_r \simeq - {\nu \over a}~~{f a^2 \over 4 p^2}~
{1 \over X^{3 \over 2} + {fa \over 2p} X}~~~~\ .
\label{vrf}
\end{equation}
Eq.~(\ref{vrf}) shows that, within the tricusp, the gas velocity 
is also inward but much larger, by a factor three or more, than 
outside the tricusp.  It increases rapidly as $r \rightarrow a_+$.
Therefore the viscous forces tend to drive the gas towards $r=a$.
This happens on gas dynamic time scales which are much shorter 
than the cosmological time scales over which the caustic rings 
migrate.  The accumulation of gas at the caustic ring radius 
may lead to enhanced star formation there.  Such processes 
may be at work in the case of the Monoceros Ring.  To the 
extent that such processes dominate the formation of the 
Monoceros Ring, the stars in the Ring should be younger
than average.  Ref. \cite{Mart06} presents evidence that 
Monoceros Ring stars are on average bluer and therefore 
younger than ordinary disk stars. 

\subsection{Effect on star orbits}

In this subsection, we consider the adiabatic deformation of 
disk star orbits by the slowly growing caustic ring of dark 
matter.  

\subsubsection{Collision rate}

Adiabatic deformation of star orbits presupposes that the star 
collision rate is small.  The time scale over which star orbits 
are significantly modified through gravitational scattering with 
other stars in a population is \cite{BT}
\begin{equation}
t_{\rm relax} \simeq 0.3~{\sigma^3 \over G^2 m^2 n \ln \Lambda}
\label{trelax}
\end{equation}
where $n$ is the density of stars in the population, $m$ 
their typical mass, $\sigma$ their velocity dispersion, and 
$\ln \Lambda \simeq 20$.  For the stars in the Monoceros Ring, 
$t_{\rm relax} \sim 10^{16}$ yr if we set $m \sim M_\odot$, 
$\sigma \sim 20$ km/s, and $n \sim 10^{-3}/{\rm pc}^3$.  Since
$t_{\rm relax}$ is much greater that the age of the universe, we
are justified in neglecting collisions among stars in the Ring.

\subsubsection{Orbit stability}

In a gravitational field $\vec{g}(\vec{r}) = - g(r) \hat{r}$, 
the angular frequency squared of small radial oscillations about 
a circular orbit of radius $r$ is 
\begin{equation}
\omega^2(r) = {1 \over r^3} {d \over dr} \left(r^3 g(r)\right)~~~\ .
\label{genom}
\end{equation}
The orbit is stable if $\omega^2 > 0$.  In the neighborhood of 
a caustic ring of dark matter, where the gravitational field is 
as given in Eq.~(\ref{fingr}), we have 
\begin{equation}
\omega^2(r) = 2 \left({v_{\rm rot} \over r}\right)^2
\left(1 + 2 f J(r-a) + r f {d J \over dr}\right)~~~\ .
\label{omc}
\end{equation}
The sum of the first two terms in the parentheses on the RHS of
Eq.~(\ref{omc}) is of order one.  The third term is
\begin{equation}
r f {v \over b} \left({d I \over dr} + {d H \over dr} \right)~~~\ .
\label{3dterm}
\end{equation}
${d I \over dr}$ is everywhere positive, whereas ${d H \over dr}$ 
has the qualitative form 
\begin{equation}
{d H \over dr} \sim - {1 \over 2 p^\prime}
{1 \over \cosh^2 ({r-a \over p^\prime})}~~~\ .
\label{dHdr}
\end{equation}
Since we expect $p^\prime$ to be of order $a$ and at any 
rate much larger than $f a \sim {1 \over 20} a$, we find 
that circular orbits are stable everywhere in the neighborhood 
of a {\it circular} caustic ring of dark matter.  The conclusion 
is valid only if the caustic ring is circular, because lack of 
axial symmetry may drive an instability through the phenomenon 
of Lindblad resonance, as we now discuss.

A particle on an orbit of radius $r$ is subjected to a radial 
time-dependent, but periodic, gravitational force from a 
non-circular caustic.  The period 
$T = {2 \pi \over \Omega} \simeq 2 \pi {r \over v_{\rm rot}}$
equals the time to go around the Galaxy once.  The equation of 
motion for the radial coordinate of the particle is that of 
a harmonic oscillator of proper frequency $\omega(r)$ driven 
by a periodic external force with frequencies 
$m \Omega(r) \simeq m {v_{\rm rot} \over r}$, where 
$m =~1,~2,~3~...$ .   Resonance occurs when $\omega(r) = 
m \Omega (r)$.  For the corresponding radii, circular orbits 
are unstable.  Using Eqs.~(\ref{crom}) and (\ref{omc}), we 
find the instability condition 
\begin{equation}
r f {d J \over dr} = ({m^2 \over 2} - 1) \left[1 + 2 f J(r-a)\right]
\label{instc}
\end{equation}
in the neighborhood of a caustic ring.  Since $2 f J(r) \ll 1$, and 
\begin{equation}
{d J \over dr} \simeq {d I \over dr} \simeq 
{1 \over 2} {1 \over \sqrt{p(r-a)}}~\Theta (r-a)~\Theta (a+p-r)
\label{dIdr}
\end{equation}
the instability occurs at radii
\begin{equation}
r_m \simeq a + f^2 {a^2 \over p} {1 \over (m^2 - 2)^2}
\label{inrad}
\end{equation}
for $m =~2,~3,~4 ...$  There is no resonance for $m=1$.  For 
$a = 20$ kpc and $f = 0.046$, the instabilities occur at 
\begin{equation}
\{r_m - a: m = 2, 3, ... \} = 
(210,~17,~...)~{\rm pc}~{{\rm kpc} \over p}
\label{rmma}
\end{equation}
Let us caution that these estimates take account only of the 
gravitational force exerted by the caustic ring of dark matter 
itself, and neglect the gravity of the baryons that have aggregated
at the caustic.

\subsubsection{Circular orbits}

In the absence of the caustic, the effective potential for radial 
motion is 
\begin{equation} 
V_{{\rm eff},0}(r) = v^2_{\rm rot} \ln r + \frac{l^2}{2r^2} 
\label{Veff}
\end{equation} 
where $l$ is the specific angular momentum of the star. When 
the caustic is present, the effective potential is 
$V_{\rm eff}(r) = V_{{\rm eff},0}(r) + V_c(r)$ where
\begin{equation} 
V_c(r) = 2 f v_{\rm rot}^2 \int \frac{dr}{r}~J(r-a)~~~~\ .
\label{Vcr}
\end{equation} 
$V_c(r)$ is plotted in Fig.~2c for
\begin{equation} 
J(r-a) = I(1,\frac{r-a}{p}) - 
\frac{1}{2} \tanh\left(\frac{r-a}{p^{'}} \right) 
\end{equation} 
with $a = 20$ kpc, $p = 1$ kpc, and $p^{'} = 5$ kpc. Fig.~2c 
illustrates the fact that the effective potential is smooth 
even though its second derivative diverges at $r=a$ and $r=a+p$. 

The caustic ring radius increases with time. According to 
the self-similar infall model, $a \propto t^{\frac{2}{3} +
\frac{2}{9\epsilon}}$.  Consider a particle of specific 
angular momentum $l$. In the absence of the caustic, it is 
on a circular orbit of radius $r$ given by
\begin{equation}
l^2 = r^3\,g(r) = r^2 \, v^2_{\rm rot}~~~~\ ,
\label{eqfr}
\end{equation}
or it is oscillating about a circular orbit of that radius. In 
the presence of a caustic of radius $a$, the particle is on 
or oscillating about a circular orbit of radius $r^{'}(r,a)$ 
given by 
\begin{equation}
l^2 = r^{'3}\,g^{'}(r^{'}) 
= r^{'2} \, v^2_{\rm rot} \left[1 + 2 f J(r^{'}-a) \right]~~~\ .
\end{equation}
Angular momentum conservation implies 
\begin{equation}
r^2 = r^{'}(r,a)^2 \, \left[1 + 2 f\, J(r^{'}(r,a) - a) \right]. 
\label{3.23}
\end{equation}
Fig.~\ref{fig3} shows $r^{'}(r,a)$ as a function of $a$. Each 
line in that figure corresponds to a different value of $r$. Let 
$\rho(r)$ be the density of stars in the absence of the caustic, 
and $\rho^{'}(r,a)$ their density in the presence of a caustic 
with radius $a$. Assuming that all stars are and remain on 
circular orbits, conservation of the number of stars implies
\begin{equation}
r^{'}\,dr^{'}\,\rho^{'}(r^{'},a) = r\,dr\,\rho(r)
\end{equation}
with $r^{'}(r,a)$ given by Eq.~(\ref{3.23}). One readily finds
\begin{equation}
\rho^{'}(r^{'},a) = \rho(r) \left[1 + 2 f\,J(r^{'}-a) + 
f r^{'}\,\frac{dJ}{dr}(r^{'}) \right]. 
\label{3.24}
\end{equation}
Assuming that the initial star density has no significant structure 
of its own, we have
\begin{equation}
\rho^{'}(r,a) = \rho \left[1 + 2 f\,J(r-a) + f r\,\frac{dJ}{dr}(r) \right]
\label{3.25}
\end{equation}
where $\rho$ is the initial density. The second term in brackets 
is an order $10\,\%$ modulation of the initial density. The third 
term is of order $f$, like the second term, but diverges when 
$r \rightarrow a_+$ as $1/\sqrt{r-a}$. It has the same singularity 
structure at $r \rightarrow a_+$ as the dark matter density because 
$g^{'}(r) = \frac{v^2_{\rm rot}}{r} \left[1 + 2 f\,J(r) \right]$
satisfies Gauss$^{'}$ law. Eq.~(\ref{3.25}) states that, in the 
limit where all stars are on circular orbits, the star density
adopts at $r=a_+$ the same divergent profile as the dark matter 
in the caustic surface there.  The pile-up of circular orbits
at $r = a_+$ is clearly seen in Fig.~\ref{fig3}.

\subsubsection{Non circular orbits}

The observed radial velocity dispersion $\Delta v$ of stars in the
Monoceros Ring is of order 20 km/s.  This implies that the stars 
do not move on circular orbits, but oscillate in the radial direction 
with typical amplitude 
\begin{equation} 
\Delta r \simeq 20~{{\rm km} \over {\rm s}}~{1 \over \omega}
\simeq 20~{{\rm km} \over {\rm s}}~{1 \over \sqrt{2}}~
{20~{\rm kpc} \over v_{\rm rot}} =~1.3~{\rm kpc}~~~~\ .
\end{equation}
The density profile of stars in the neighborhood of a caustic
ring is therefore the profile of Eq.~(\ref{3.24}) averaged over
the length scale $\Delta r$. The sharp features at $r=a$ and
$r = a+p$ are then washed out. However, there will be an 
average relative overdensity within the tricusp
\begin{equation}
\left< \frac{\rho^{'} - \rho}{\rho} \right> \simeq
af \left< \frac{dJ}{dr} \right >\bigg| _ {a< r< a+p}
\sim \frac{af}{p} \simeq 1~{{\rm kpc} \over p}
\end{equation}
provided $p$ is not much less than $\Delta r$. This is the 
$100 \%$ average overdensity of ordinary disk stars at the 
$n=2$ caustic ring of dark matter, announced earlier.

For the proposed interpretation of the Monoceros Ring, the 
radial velocities acquired by stars as a result of the passing 
of a caustic ring should be less than the measured radial velocity 
dispersion of stars in the Monoceros Ring.  We now verify that 
this is the case. 

The radial motion of a star in a near circular orbit in the 
neighborhood of a caustic ring is that of a harmonic oscillator 
whose equilibrium position is $r^{'}\left(r(l),a(t)\right)$ and 
whose proper frequency $\omega$ is given by Eq.~(\ref{omc}) 
evaluated at $r^{'}$.  $a(t)$ is a smooth function of time. 
However, as shown in Fig.~\ref{fig3}, $r^{'}$ as a function 
of $a$ has kinks at $a=r^{'}$ and $a = r^{'} - p$.  The 
velocity $\frac{dr^{'}}{dt}$ of the equilibrium position 
changes abruptly when it passes the caustic at those locations.
A star in circular orbit in the $z=0$ plane first encounters 
the caustic at $r^{'} = a+p$.  For illustrative purposes, we 
consider the case $a = 20$ kpc, $t_0 = 13.7$ Gyr, $\epsilon = 0.3$, 
$p = 1$ kpc, $v=b$ and $f = 0.046$. The caustic moves with outward 
speed 
\begin{equation}
\frac{da}{dt} = 
\left(\frac{2}{3} + \frac{2}{9\epsilon}\right ) \frac{a}{t_0} 
= 2.0~{\rm km/s}~~~~\ . 
\end{equation}
When $r^\prime = a+p$, the velocity of the equilibrium position 
changes abruptly by the amount ($\delta$ is infinitesimally small)
\begin{eqnarray}
\Delta~\frac{dr^{'}}{dt} 
&=& \left[ \frac{dr^{'}}{da}\left(r^{'} = a + p - \delta \right) 
- \frac{dr^{'}}{da}\left( r^{'} = 
a + p + \delta \right) \right] \frac{da}{dt} \nonumber \\
&\simeq& \frac{af}{2p + af} \frac{da}{dt}~\simeq~0.6~{\rm km/s}~~~\ .
\end{eqnarray}
Here and below we are neglecting terms of order $f$, $p/a$, 
$p/p^\prime$ and $af/p^\prime$ versus terms of order one.  
The proper oscillation frequency after the caustic has passed is
\begin{equation}
\omega(r = a + p - \delta) \simeq 
\sqrt{2}~\frac{v_{rot}}{a} \sqrt{ 1 + \frac{af}{2p}} 
\simeq \frac{1}{5 \times 10^7~\rm{years}}~~~\ .
\end{equation}
Thus a star which is in a circular orbit in the $z=0$ plane before 
the $r^{'} = a+p$ caustic passes by, will be oscillating in the 
radial direction with initial amplitude
\begin{equation}
A = \frac{1}{\omega} \Delta~\frac{dr^{'}}{dt} \simeq 30~{\rm pc} 
\label{3.28}
\end{equation}
after that caustic passes by.  Stars not in the $z=0$ plane do not 
go through the caustic cusp at $r^{'}=a+p$, and presumably make
smoother transitions through the caustic than the stars in the 
$z=0$ plane. So, Eq.~(\ref{3.28}) provides an upper limit for the 
$z \ne 0$ stars. Stars receive a second jolt when the $r^{'} = a$
caustic surface passes by. The analogous quantities for that 
transition are :
\begin{eqnarray}
\Delta~\frac{dr^{'}}{dt} &=& \left[\frac{dr^{'}}{da}(r^{'}= a_-) 
- \frac{dr^{'}}{da}(r^{'}= a_+) \right] \frac{da}{dt} 
\simeq - \frac{da}{dt} \simeq -2~{\rm km/s}\nonumber \\
\omega(r = a_-) &\simeq& \sqrt{2}~\frac{v_{\rm rot}}{a} \simeq
\frac{1}{6 \times 10^7~{\rm year}}\nonumber \\
A &=& \bigg|\Delta~{dr^{'} \over dt} \bigg| \frac{1}{\omega} 
\simeq 130~{\rm pc}~~~\ .
\end{eqnarray}
After the equilibrium position $r^{'}$ has passed by the caustic 
surface at $r^{'}=a$, $r^{'}$ decreases very quickly, implying a 
depletion on the $r < a$ side of the caustic. This depletion is
evident in Fig.~\ref{fig3}.

\section{Conclusions}
 
The flow of cold collisionless dark matter in and out of
the gravitational potential of a galaxy necessarily forms
inner and outer caustics.  The inner caustics are rings in
the galactic plane provided the angular momentum distribution
of the infalling dark matter is characterized by net overall
rotation.  Assuming self-similar infall, the radii of the 
caustic rings of dark matter in our galaxy were predicted
to be  40 kpc/$n$  with $n$ = 1, 2, 3 ...  Because the Monoceros
Ring of stars is located near the second caustic ring of dark 
matter we looked for processes by which the latter may cause 
the former.

We have identified two such processes.  The first is the flow 
of gas in the disk towards the sharp angular velocity minimum 
located at the caustic ring radius, increasing the rate of star 
formation there.  To the extent that this process is responsible
for the formation of the Monoceros Ring, the Ring stars are 
predicted to be bluer than average. The second process is the 
adiabatic deformation of star orbits in the neighborhood of the 
caustic ring.  As the spatial dependence of the gravitational field 
of a caustic ring is known, it is straightforward to obtain the 
map of initial to final orbits for disk stars.  The resulting 
enhancement of disk star density at the location of the second 
caustic ring is of order 100\%.  Because of uncertainties in the 
caustic parameters ${v \over b}$ and $p$, and in the velocity 
distribution of the disk stars, the strength of the enhancement 
can only be estimated within a factor of two or so.

The self similar infall model of galactic halo formation is expected 
to describe the halos of all isolated spiral galaxies.  The caustic 
rings of dark matter in exterior galaxies may also be revealed by the
baryonic matter they attract.  Our analyis is relevant to those cases 
as well.

The existence of caustics has implications for most approaches 
to the detection of dark matter, including direct searches of 
WIMPs \cite{WIMP} and axions \cite{axion}, the gamma ray signal 
from WIMP annihilation \cite{gamma}, and gravitational lensing
\cite{lens}.

\acknowledgments

We thank Scott Tremaine for extended discussions and for 
several insights which played a crucial role in the development 
of the results presented here.  This work was supported in part 
by the U.S. Department of Energy under grant DE-FG02-97ER41209.  
P.S. gratefully acknowledges the hospitality of the Aspen Center 
for Physics while he was working on this project.


\begin{figure}[]

\subfigure[$\;$]{\label{fig1-a}
\includegraphics[width=3.5in,height=3.5in]{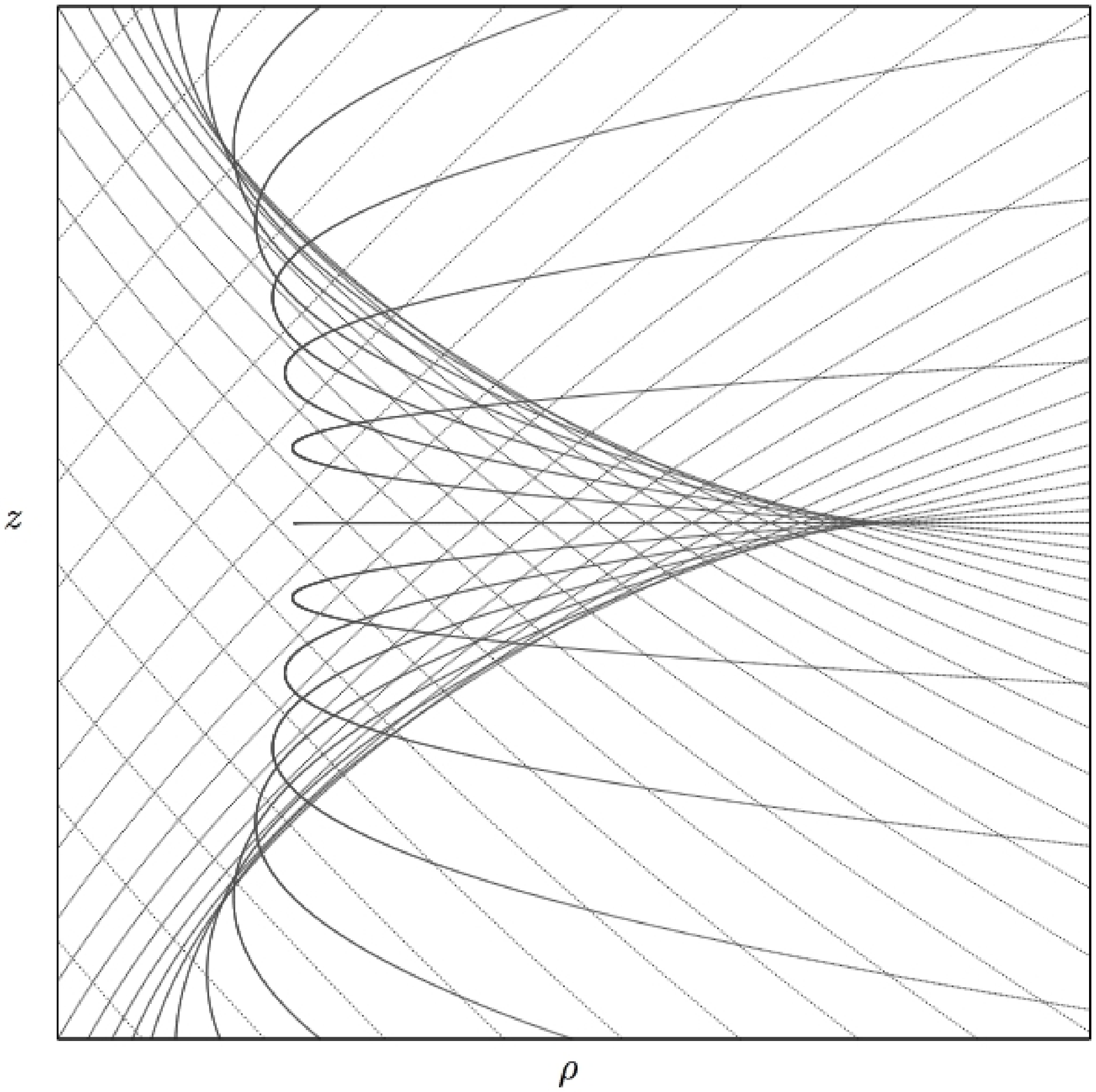}} 
\subfigure[$\;$]{\label{fig1-b}
\includegraphics[width=3.5in,height=3.5in]{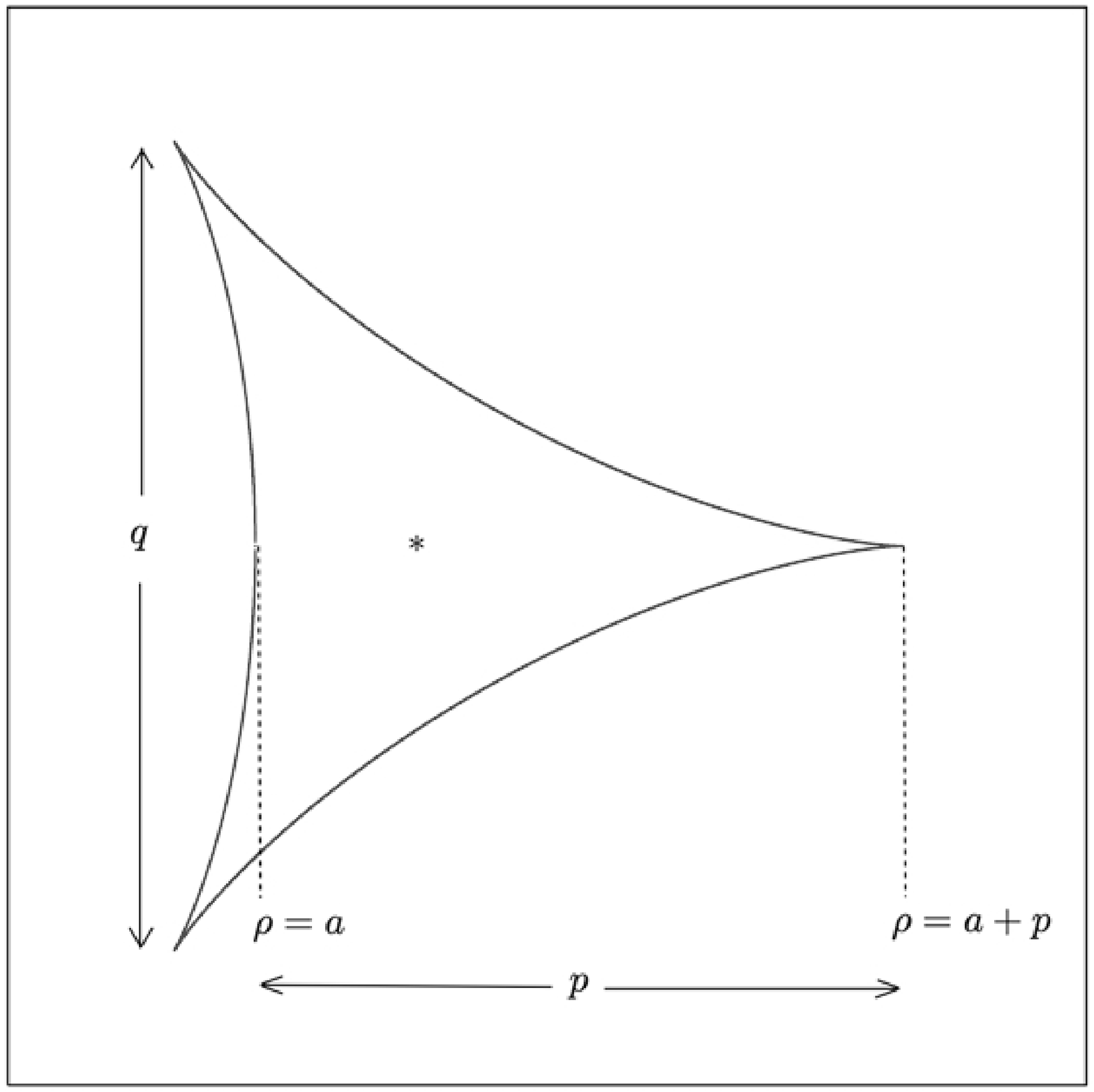}}
\caption{(a) Dark matter trajectories forming a caustic ring of 
dark matter, in $\rho-z$ cross-section.  (b) The envelope of 
the trajectories shown in (a).  We refer to the shape shown in (b) 
as the ``tricusp".  The figure also indicates what is meant by 
the radius $a$ of a caustic ring, and by its transverse sizes $p$
and $q$.  The star indicates the center of the tricusp, as defined 
in the text.\label{fig1}}
\end{figure}

\begin{figure}[]

\subfigure[$\;\;$]{\label{fig2-a}
\includegraphics[width=4in,height=2in]{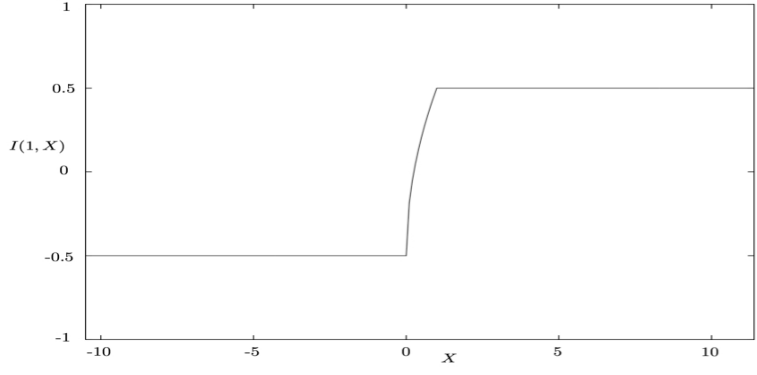}}
\subfigure[$\;\;$]{\label{fig2-b}
\includegraphics[width=4in,height=2in]{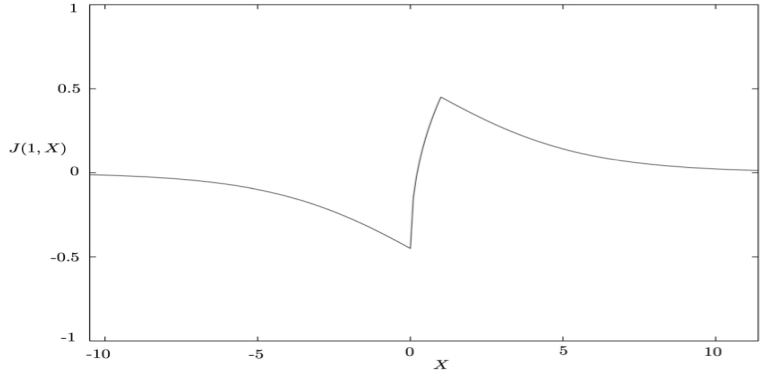}} 
\subfigure[$\;\;$]{\label{fig2-c}
\includegraphics[width=4in,height=2in]{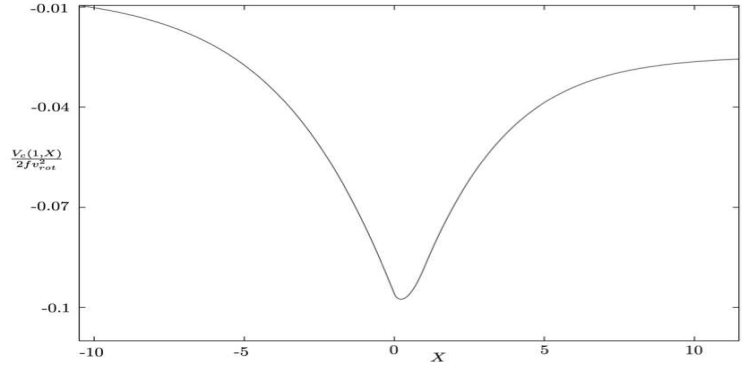}}
\caption{(a) The function $I(1,X)$ defined in Eq.~(\ref{I1X}), 
with $X \equiv {r-a \over p}$. (b) The function $J(X)$ defined 
in Eq.~(\ref{Jr}) for $v=b$, $\zeta=1$ and 
$H(X) = - {1 \over 2} \tanh({X \over 5})$. (c) The function $V_c(X)$ 
defined in Eq.~(\ref{Vcr}), in units of $2 f v_{\rm rot}^2$, for 
$J(X)$ as in (b). \label{fig2}} \end{figure}

\begin{figure} 
\resizebox{4in}{4in}{\includegraphics{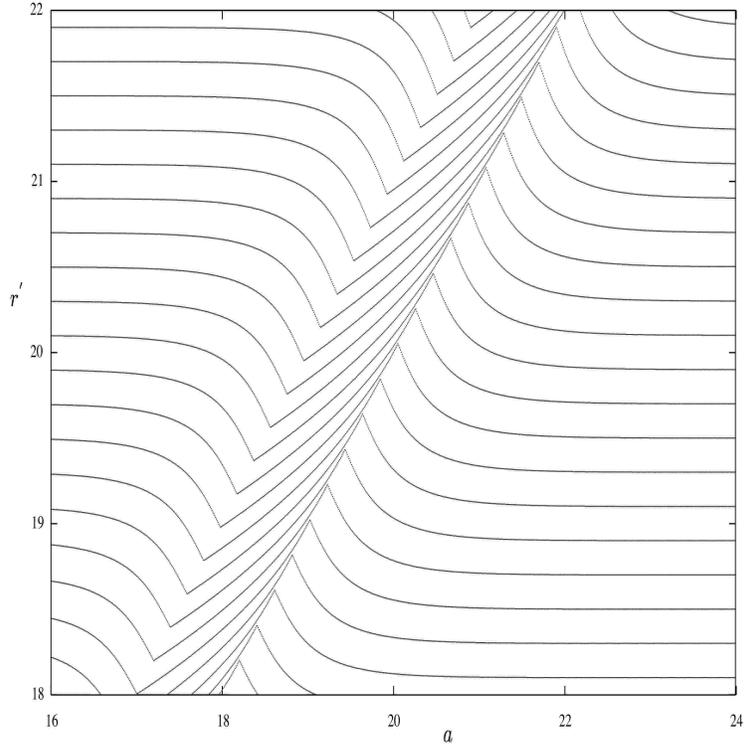}} 
\caption{The radii $r^\prime(r,a)$ of circular orbits in the presence 
of a caustic, as a function of the caustic radius $a$.  $r$ is the
radius of the orbit in the absence of caustic.  Each line corresponds 
to a different value of $r$.  Each line has an upward kink at 
$r^\prime = a + p$ and a downward kink at $r^\prime = a$.  As $a$
increases with $p$ fixed, $r^\prime$ first decreases from its 
initial value $r$ until $r^\prime = a+p$.  For $a < r^\prime < a+p$,
i.e. when the orbit is within the tricusp, $r^\prime$ increases with 
$a$ but not as fast as $a$, until $a = r^\prime$.  As $a$ increases 
yet further $r^\prime(r,a)$ returns to the value $r$ from which it
started. Note that the radii of circular orbits pile up at $r=a$.  
Stars on approximately circular orbits oscillate about $r^\prime(r,a)$.}
\label{fig3} 
\end{figure}

\end{document}